\documentclass[entropy,article,submit,moreauthors,pdftex]{Definitions/mdpi} 
\usepackage[percent]{overpic}
\usepackage{verbatim}

\firstpage{1} 
\pubvolume{xx}
\issuenum{x}
\articlenumber{xx} 
\pubyear{2020}
\copyrightyear{2020}
\history{Received: date; Accepted: date; Published: date}

\Title{Electronic wave-packets in integer quantum Hall edge channels: relaxation and dissipative effects}

\Author{Giacomo Rebora$^{1,2}$, Dario Ferraro$^{1,2,*}$, Ramiro H. Rodriguez$^{3}$, François D. Parmentier$^{3}$, Patrice Roche$^{3}$ and Maura Sassetti$^{1,2}$ }
 
\address{
$^{1}$ \quad Dipartimento di Fisica, Universit\`a di Genova, Via Dodecaneso 33, 16146, Genova, Italy\\
$^{2}$ \quad SPIN-CNR, Via Dodecaneso 33, 16146, Genova, Italy\\
$^{3}$ \quad Universit\'e Paris-Saclay, CEA, CNRS, SPEC, Gif-sur-Yvette 91191, France.
}

\corres{Correspondence: ferraro@fisica.unige.it} 

\preto{\abstractkeywords}{\nolinenumbers}

\abstract{
We theoretically investigate the evolution of the peak height of an energy resolved electronic wave-packets ballistically propagating along integer quantum Hall edge channels at filling factor $\nu=2$. This is ultimately related to the elastic scattering amplitude for the fermionic excitations evaluated at different injection energy. We investigate this quantity assuming a short range capacitive coupling between the edges. Moreover, we also take into account phenomenologically the possibility of energy dissipation towards additional degrees of freedom both linear and quadratic in the injection energy. Comparing with recent experimental data, we rule out the non-dissipative case as well a quadratic dependence of the dissipation, indicating a linear energy loss rate as the best candidate to describe the behavior of the quasi-particle peak at short enough propagation lengths.}

\keyword{Electron quantum optics; interaction effects; relaxation; dissipation}

\begin{document}
\section{Introduction}
The possibility to prepare, manipulate and measure individual electronic wave-packets propagating along mesoscopic quantum channels such as the edge states of the quantum Hall (QH) liquid opened the way to the so called Electron Quantum Optics (EQO) \cite{Feve07, Grenier11, Bocquillon14, Roussel17, Ferraro17b, Glattli17, Bauerle18}. In this framework, seminal quantum optics experiments such as the Hanbury-Brown-Twiss \cite{Hanbury56} and the Hong-Ou-Mandel \cite{Hong87} interferometry have been realized by using ballistic electrons \cite{Bocquillon12, Bocquillon13b, Dubois13b, Freulon15, Marguerite16}. 

The extremely high level of control reached in this domain triggered an intense interest on the possibility to use electronic excitations as flying qubits \cite{Yamamoto12, Bautze14, Gaury14, Bauerle18}, namely as a controlled and trustful way to transport quantum information over relatively long distances \cite{DiVincenzo00}. However, one of the main challenges to implement this idea in realistic solid state devices is represented by the interaction among the electrons in the system and with the external environment, effects with no parallel in the photonic case. The role of Coulomb interaction has been extensively discussed both in order to properly understand the experimental observation achieved for integer states at filling factor $\nu=2$ \cite{Marguerite16, Wahl14, Ferraro14, Cabart18, Acciai18, Rebora20} and to predict new features occurring in the strongly interacting fractional QH regime \cite{Ferraro15, Rech17, Vannucci17, Ronetti18, Ferraro18b, Ronetti19} or in more exotic low dimensional systems \cite{Ziani15, Cavaliere16, Dolcetto16}. Conversely, the role of energy dissipation towards external degrees of freedom on individual electrons is still largely unexplored. 

Preliminary steps along this direction involved the study of the evolution of a non-equilibrium electronic distribution as a function of the interaction length \cite{Altimiras10, leSueur10}, where experiments showed signature of important energy losses towards external degrees of freedom, included only effectively in the theoretical models \cite{Degiovanni10, Lunde10, Kovrizhin12}. Such dissipation effects are crucial in order to properly describe both the dynamics of integer QH states \cite{Bocquillon13} and the evolution of the peak height of energy resolved wave-packets injected into them \cite{Rodriguez20}. Remarkably enough, the predicted functional form of the dissipation as a function of the energy is different in these two cases, namely quadratic in the former case, linear in the latter. Therefore, a more detailed analysis is needed in order to clarify this apparent discrepancy and to improve our understanding on this topic. 

The present paper aim at addressing this subject. We start from a hydrodynamic model, where the two edge channels are capacitively coupled through a short range interaction \cite{Wen95}. In addition, we consider three possible dissipation regimes: the non-dissipative case used as reference case, an ohmic dissipation linear in the injection energy of the electronic wave-packet and a quadratic dissipation. We observe that the linear dependence provides the best fit for the experimental data of the evolution of the experimental peak height at small enough propagation length~\cite{Rodriguez20}. Conversely, at greater propagation lengths, a dissipation quadratic in the injection energy dominates~\cite{Bocquillon13}. This apparent discrepancy could be related to both different sample design or more involved functional dependence of the dissipation.

The paper is organized as follows. In Section \ref{sec:model} we discuss the edge state at $\nu=2$, where the two channels are capacitively coupled, in terms of a bosonic hydrodynamic model. Section \ref{sec:scattering} describes the edge-magnetoplasmon scattering matrix connecting the bosonic fields incoming into the interacting region with the outgoing ones. Here, we include also the effects of energy dissipation towards external degrees of freedom. In particular, we consider: a non-dissipative case, dissipation with a linear and a quadratic dependence on the injection energy. The elastic scattering probabilities for the fermionic excitations in the various regimes is reported in Section \ref{sec:elastic} and the comparison with experimental data is given is Section \ref{sec:result}. Section \ref{sec:conclusions} is devoted to the conclusion, while we have included technical details of the derivation of the elastic scattering amplitude in Appendix \ref{AppA}.


\section{Model \label{sec:model}}
We consider the two edge channels of a QH bar at filling factor $\nu=2,$ assuming a short range ($\delta$-like) capacitive coupling between them. Considering the conventional Wen's hydrodynamical approach \cite{Wen95} for this system one can write the Hamiltonian density ($\hbar=1$) \cite{Braggio12}
\begin{equation}
\mathcal{H}{(x)}= \frac{v_{1}}{4 \pi} \left(\partial_{x} \phi_{1}(x) \right)^{2}+\frac{v_{2}}{4 \pi} \left(\partial_{x} \phi_{2}(x) \right)^{2}+\frac{u}{2 \pi} \partial_{x} \phi_{1}(x) \partial_{x} \phi_{2}(x)
\label{Hamiltonian}
\end{equation}
where $\phi_{i}$ ($i=1,2$) are bosonic fields related to the $i$-th edge particle density through the condition  
\begin{equation}
\rho_{i}(x)= \frac{1}{2 \pi} \partial_{x} \phi_{i}(x),
\end{equation} 
$v_{i}$ are the bare propagation velocities of the bosonic modes along the two channels and $u$ the intensity of their coupling. Without loss of generality, in the following we will assume $v_{1}\geq v_{2}$.

The above Hamiltonian can be diagonalized by means of a rotation in the bosonic field space of the form 
\begin{equation}
\mathcal{R}\left( \theta\right)=\left(\begin{matrix} 
\cos \theta & \sin \theta \\
-\sin\theta & \cos \theta 
\end{matrix}\right)
\end{equation}
with a mixing angle satisfying 
\begin{equation}
\tan \left( 2 \theta \right)= \frac{2 u}{v_{1}-v_{2}}.
\end{equation}

The above condition naturally leads to two different regimes. 

\subsection{"Strongly interacting" regime}
The condition typically indicated in literature as "strongly interacting" is characterized by
\begin{equation}
\theta=\frac{\pi}{4}
\end{equation}
which can be actually achieved only in the symmetric case $v_{1}=v_{2}=v$. This limit is usually assumed as work hypothesis in the majority of the theoretical papers \cite{Levkivskyi08, Degiovanni10, Kovrizhin12, Wahl14, Ferraro14, Slobodeniuk16, Ferraro17, Acciai18} and leads eigenvelocities 
\begin{equation}
v_{\rho, \sigma}= v\pm u
\end{equation}
where $v_\rho$ is associated to a charge eigenmode $\phi_\rho$, while $v_\sigma$ corresponds to a dipole eigenmode $\phi_\sigma$.
It is worth noticing that the stability condition of the model, namely the fact that both eigenvelocities need to be positive, leads the further constraint $v> u$. This implies that the coupling between the channels cannot be arbitrary high, contradicting the conventional denomination.

Even if frequently used in order to fit experimental data \cite{Degiovanni10, Bocquillon13, Marguerite16} this approximation revealed too restrictive in some cases \cite{Tewari16}. 

\subsection{"Moderately interacting" regime} 
\begin{figure}
\centering
\includegraphics[scale=0.7]{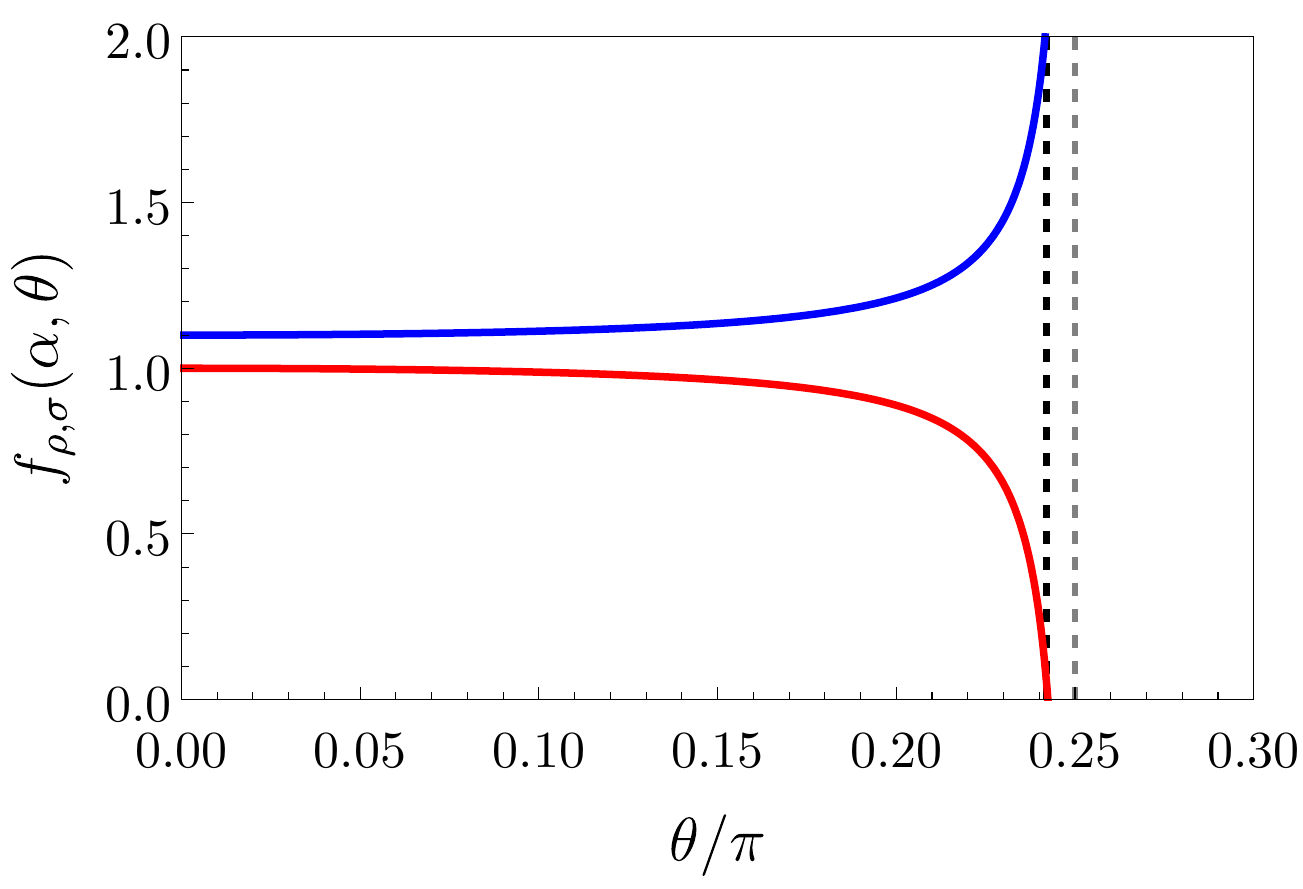}
\caption{Plot of $f_{\rho}(\alpha, \theta)$ (full blue curve) and of $f_{\sigma}(\alpha, \theta)$ (full red curve) as a function of $\theta$ (in units of $\pi$) for fixed $\alpha=1.1$. Vertical dashed lines are placed at $\theta=\frac{1}{2} \arccos\left(\frac{\alpha-1}{\alpha+1} \right)\approx 0.242 \pi$ ("moderately interacting" regime in blue) and $\theta=\pi/4$ ("strongly interacting" regime in gray) as references.}
\label{Fig1}
\end{figure}

In order to relax the above constraints one can assume, without loss of generality, $v_{2}=v$ and $v_{1}= \alpha v$, with $\alpha>1$. Notice that for $\alpha=1$ we recover the previous case. Under these conditions, the two eigenvelocities of the model become 
\begin{equation}
v_{\rho, \sigma}= v f_{\rho, \sigma}\left(\alpha, \theta\right)
\end{equation}
with 
\begin{equation}
f_{\rho, \sigma}\left(\alpha, \theta \right)= \left(\frac{\alpha+1}{2} \right)\pm \frac{1}{\cos\left(2 \theta \right)} \left(\frac{\alpha-1}{2} \right).
\end{equation}
The stability condition of the model \cite{Braggio12} imposes the constraint 
\begin{equation}
\theta\leq \frac{1}{2} \arccos\left(\frac{\alpha-1}{\alpha+1} \right) <\frac{\pi}{4},
\end{equation}
manifestly more restrictive with respect to the "strongly interacting" case ($\alpha=1$). The behaviors of $f_{\rho}$ and $f_{\sigma}$ as a function of $\theta$ and at fixed $\alpha$ are shown in Fig.~\ref{Fig1}. In the following, we will focus on this general case which seems more realistic in order to properly describe experimental observations. 

\section{Edge-magnetoplasmon scattering matrix \label{sec:scattering}}
\begin{figure}[h]
\centering
\includegraphics[scale=0.55]{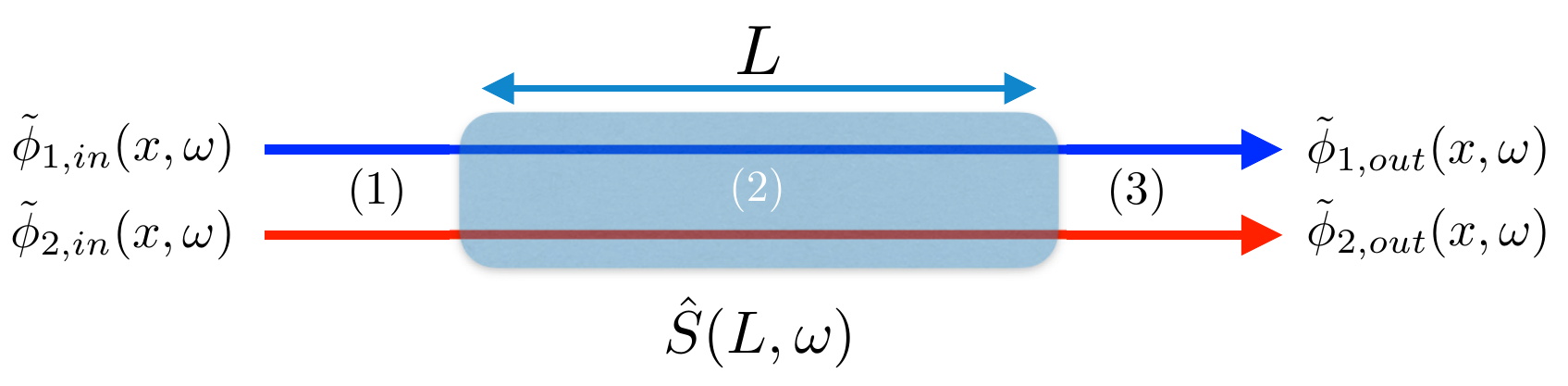}
\caption{Model for a QH edge state at filling factor $\nu=2$. According to the chirality one can identify the incoming (injection) region $(1)$, the interacting region $(2)$ (shaded area of length $L$) and the outgoing (detection) region $(3)$. In regions $(1)$ and $(3)$ the dynamics of the bosonic fields is well described in terms of free equations of motion ($u=0$). Moreover, the outgoing fields, written in the Fourier space ($\tilde{\phi}_{i,out}(x,\omega)$, with $i=1,2$), are connected to the incoming ones ($\tilde{\phi}_{i,in}(x,\omega)$, with $i=1,2$) through the edge-magnetoplasmon scattering matrix $\hat{S}(L, \omega)$ which encodes the information of the inter-channel interaction acting over a length $L$ and at a given frequency (energy) $\omega$.}
\label{Fig2}
\end{figure}

The experiment discussed in Ref. \cite{Rodriguez20} involves the injection of an electronic wave-packet with Lorentzian profile in energy and its detection after a given propagation length along the edge. In order to describe this situation one can proceed as in Refs. \cite{Degiovanni10, Ferraro14, Ferraro17}, where the edge channels are divided into three parts: a non-interacting injection region, an interacting propagating region and a non-interacting region of detection (see Fig. \ref{Fig2}). Notice that this separation is not an oversimplification of the problem. Indeed, chirality guarantees that the interacting region can be made arbitrary close both to the injection and the detection regions without loss of generality. In the following we will discuss in detail the dynamics of the edge channels in the various regions. 
\begin{itemize}

\item \emph{Injection region $(1)$}

In this region one can assume $u=0$ and the Hamiltonian density can be simply written as 
\begin{equation}
\mathcal{H}^{(1)}(x)= \frac{v_{1}}{4 \pi} \left(\partial_{x} \phi_{1,in}(x)  \right)^{2}+\frac{v_{2}}{4 \pi} \left(\partial_{x} \phi_{2,in}(x)  \right)^{2}.
\end{equation}
The bosonic fields $\phi_{1,in}$ and $\phi_{2,in}$ propagate freely according to the equations of motion
\begin{equation}
\left(\partial_{t} +v_{i}\partial_{x}\right)  \phi_{i,in}(x, t) =0.
\label{Region1}
\end{equation}
By moving into Fourier transform with respect to time they become 
\begin{equation}
\left(-i \omega +v_{i} \partial_{x}\right) \tilde{\phi}_{i,in}(x, \omega) =0,
\end{equation}
with $\tilde{\phi}_{i,in}(x, \omega)$ field amplitudes in the frequency space defined as
\begin{equation}
    \tilde{\phi}_i(x,\omega)=\int  e^{i\omega t} \phi_i(x,t)\,d\omega.
\end{equation}\\
\item \emph{Interacting region $(2)$}

In this region the Hamiltonian density is the one in Eq. (\ref{Hamiltonian}). According to the previous discussion, the bosonic fields $\phi_{1}$ and $\phi_{2}$ are no longer eigenstates of the Hamiltonian and the system is diagonalized in terms of a charged and a dipole mode, indicated respectively with $\phi_{\rho}$ and $\phi_{\sigma}$ with the eigenvelocities $v_{\rho}$ and $v_{\sigma}$ as discussed above. In this case the equations of motion are
\begin{equation}
\left(\partial_{t}+v_{\eta}\partial_{x}\right) \phi_{\eta}(x, t) = 0 \qquad \eta= \rho, \sigma
\end{equation}
that, expressed in Fourier transform with respect to time, become 
\begin{equation}
\left(-i \omega +v_{\eta} \partial_{x}\right) \tilde{\phi}_{\eta}(x, \omega) =0.
\label{EOM_interaction}
\end{equation}

The solution of the equations of motion in this region reads 

\begin{equation}
\tilde{\phi}_{\eta} (x, \omega)= e^{i \frac{\omega}{v_{\eta}} x} \tilde{\phi}_{\eta} (0, \omega)
\label{EOM_solution}
\end{equation}
with 
\begin{eqnarray}
\tilde{\phi}_{\rho}(0, \omega)&=& \cos\theta\tilde{\phi}_{1,in}(0, \omega)+\sin \theta \tilde{\phi}_{2,in}(0, \omega) \nonumber \\ 
\tilde{\phi}_{\sigma}(0, \omega)&=& -\sin\theta\tilde{\phi}_{1,in}(0, \omega)+\cos \theta \tilde{\phi}_{2,in}(0, \omega)
\end{eqnarray}
the (possibly frequency dependent) amplitudes achieved by imposing the continuity of the fields at $x=0$ (boundary between regions $(1)$ and $(2)$). \\
\item \emph{Detection region $(3)$}

Analogously to what discussed for region $(1)$, also in this case inter-channel interaction is negligible and the equations of motion write as in Eq. (\ref{Region1}) $(\mathcal{H}^{(1)}=\mathcal{H}^{(3)})$.
Here, imposing the continuity of the fields at $x=L$ (boundary between regions $(2)$ and $(3)$), we observe that the outgoing field amplitudes are related to the incoming ones through the relations 
\begin{eqnarray}
\tilde{\phi}_{1,out}(L, \omega)&=& \cos\theta\tilde{\phi}_{\rho}(L, \omega)-\sin \theta \tilde{\phi}_{\sigma}(L, \omega) \nonumber \\ 
\tilde{\phi}_{2,out}(L, \omega)&=& \sin\theta\tilde{\phi}_{\rho}(0, \omega)+\cos \theta \tilde{\phi}_{\sigma}(L, \omega).
\end{eqnarray}
\end{itemize}


\subsection{Dissipative effects}
Experimental observations \cite{Bocquillon13, leSueur10, Rodriguez20} suggest a relevant role played by energy dissipation towards additional degrees of freedom in the transport along QH edge channels. The simplest way to include this effect is by adding a real frequency dependent energy loss rate $\gamma \left(\omega\right)$ (assumed here equal for both channels for sake of simplicity) at the level of the equations of motion in the interacting region (see Eq. (\ref{EOM_interaction})). According to this, they become
\begin{equation}
\left[-i \omega +\gamma\left(\omega\right)+v_{\eta} \partial_{x}\right] \tilde{\phi}_{\eta}(x, \omega) =0.
\label{EOM_dissipation}
\end{equation}
In the following we will focus on three possible behaviors for $\gamma(\omega)$: a non-dissipative case $\gamma=0$, a linear dissipation case $\gamma\left(\omega \right)=\gamma_{1} \omega$ ($\gamma_1$ real adimensional parameter)~\cite{Braggio12} and a dissipation quadratic in the energy $\gamma(\omega)=\gamma_{2} \omega^{2}$ ($\gamma_2$ real parameter with the dimension of a time)~\cite{Bocquillon13}.

Due to this additional contribution, the solution of the equations of motion in Eq. (\ref{EOM_solution}) acquire a frequency dependent damping 
\begin{equation}
\tilde{\phi}_{\eta} (x, \omega)= e^{i\left[\omega+i \gamma(\omega) \right] \frac{x}{v_{\eta}}} \tilde{\phi}_{\eta} (0, \omega).
\end{equation}


\subsection{General form of the scattering matrix}
According to the previous considerations and proceeding as in Ref. \cite{Ferraro17} we obtain the edge-magnetoplasmon scattering matrix connecting the incoming (injected) and the outgoing (detected) bosonic fields, namely
\begin{equation}
\left( 
\begin{matrix}
\tilde{\phi}_{1,out}(L, \omega)\\
\tilde{\phi}_{2,out}(L, \omega)\\
\end{matrix}
\right)=
\hat{S}(L, \omega)
\left( 
\begin{matrix}
\tilde{\phi}_{1,in}(0, \omega)\\
\tilde{\phi}_{2,in}(0, \omega)\\
\end{matrix}
\right), 
\end{equation}
with 
\begin{equation}
\hat{S}(L, \omega)=
\left( 
\begin{matrix}
\cos^{2}\theta e^{i \left[\omega+i\gamma(\omega)\right] \tau_{\rho}}+\sin^{2}\theta e^{i  \left[\omega+i\gamma(\omega)\right] \tau_{\sigma}} &  \sin \theta \cos \theta \left(e^{i  \left[\omega+i\gamma(\omega)\right] \tau_{\rho}}-e^{i \left[\omega+i\gamma(\omega)\right] \tau_{\sigma}} \right)\\
\sin \theta \cos \theta \left(e^{i  \left[\omega+i\gamma(\omega)\right] \tau_{\rho}}-e^{i \left[\omega+i\gamma(\omega)\right] \tau_{\sigma}} \right)&  \sin^{2}\theta e^{i \left[\omega+i\gamma(\omega)\right] \tau_{\rho}}+\cos^{2}\theta e^{i \left[\omega+i\gamma(\omega)\right] \tau_{\sigma}}\\
\end{matrix}
\right).
\label{S_matrix}
\end{equation}
In the above equation, we have introduced the short-hand notation $\tau_{\rho, \sigma}= L/v_{\rho, \sigma}$ for the times of flight associated to the propagation velocity of the charge and dipolar eigenmodes along the interacting region. 

In the following, we will focus only on the top left entry of the scattering matrix in Eq. (\ref{S_matrix}), which represents the amplitude probability for the edge-magnetoplasmon to be transmitted along the first channel (assumed as the injection/detection channel), namely 
\begin{eqnarray}
t\left(\omega\right)&=&\cos^{2}\theta e^{i \left[\omega+i\gamma(\omega)\right] \tau_{\rho}}+\sin^{2}\theta e^{i \left[\omega+i\gamma(\omega)\right] \tau_{\sigma}}\\
&=& p_{\rho}\left(\theta \right) e^{i \left[\omega+i\gamma(\omega)\right] \tau_{\rho}}+p_{\sigma}\left(\theta \right) e^{i \left[\omega+i\gamma(\omega)\right] \tau_{\sigma}}.
\label{t_omega}
\end{eqnarray} 


\section{Elastic scattering amplitude \label{sec:elastic}}
As discussed in Ref. \cite{Degiovanni09}, assuming a very peaked (ideally $\delta$-like) injected wave-packet in energy, the relative height of this peak as a function of the energy is given, at zero temperature, by 
\begin{equation}
\mathcal{V}\left(\varepsilon\right)=\frac{|\mathcal{Z}\left( \varepsilon \right)|^{2}}{|\mathcal{Z}\left(0\right)|^{2}}
\end{equation}
with 
\begin{equation}
\mathcal{Z}\left(\varepsilon \right)= \int_{-\infty}^{+\infty}d\tau e^{i \varepsilon \tau} \exp{\left\{\int_{0}^{+\infty}\frac{d \omega}{\omega} \left[t \left( \omega \right) e^{-i \omega \tau}-1 \right]e^{-\omega/\omega_{c}} \right\}}
\end{equation}
the elastic scattering amplitude (see Appendix \ref{AppA} for more details of the calculation). Here, we introduced a converging factor $\omega_{c}$ corresponding to the greatest energy scale in the systems and that will be send to $\omega_{c}\rightarrow +\infty$ at the end of the calculation \cite{Ferraro10}. Notice that this picture can be used also to describe more realistic wave-packets in the energy domain as long as their width (energy dispersion) is not too big with respect to the average energy injection, condition which is typically fulfilled in experiments \cite{Bocquillon13b, Marguerite16, Rodriguez20}.

In the following we will consider the behavior of $\mathcal{V}$ as a function of the energy for the various possible dissipations.

\subsection{Non-dissipative case} 
In absence of energy losses towards external degrees of freedom the edge-magnetoplasmon transmission amplitude is  

\begin{equation}
t_{nd}\left(\omega\right)=p_{\rho} e^{i \omega \tau_{\rho}}+p_{\sigma} e^{i \omega \tau_{\sigma}}.
\end{equation}
This leads, in the time domain, to  

\begin{eqnarray}
\mathcal{Z}_{nd}(t)&=& \exp\left\{p_{\rho} \int^{+\infty}_{0} \frac{d \omega}{\omega} \left[e^{-i \omega( t-\tau_{\rho})}-1 \right]e^{-\omega/\omega_{c}} \right\}
\exp\left\{p_{\sigma}\int^{+\infty}_{0} \frac{d \omega}{\omega} \left[e^{-i \omega( t-\tau_{\sigma})}-1 \right] e^{-\omega/\omega_{c}} \right\}
\nonumber\\
&=& \frac{-i}{\omega_{c}}\frac{1}{\left(t-\tau_{\rho}-\frac{i}{\omega_{c}}\right)^{p_{\rho}}  \left(t-\tau_{\sigma}-\frac{i}{\omega_{c}}\right)^{p_{\sigma}} }.
\end{eqnarray}

Its Fourier transform reads 

\begin{eqnarray}
\mathcal{Z}_{nd}(\varepsilon)&=& \frac{-i}{\omega_{c}}\int^{+\infty}_{-\infty} dt \frac{e^{i\varepsilon t}}{\left(t-\tau_{\rho}-\frac{i}{\omega_{c}}\right)^{p_{\rho}}  \left(t-\tau_{\sigma}-\frac{i}{\omega_{c}}\right)^{p_{\sigma}} }\nonumber\\
&=&\frac{2\pi}{\omega_{c}} e^{i \frac{\varepsilon}{\varepsilon_{0} f_{\rho}}} {}_1 F_{1} \left[ p_{\rho} , 1; -i\frac{\varepsilon}{\varepsilon_{0}} \left(\frac{1}{f_{\sigma}}-\frac{1}{f_{\rho}}\right)\right]\Theta(\varepsilon)
\end{eqnarray}
with 
\begin{equation}
\varepsilon_{0}= \frac{v}{L}, 
\end{equation}
$\Theta(...)$ the Heaviside Theta function and where ${}_1 F_{1}$ indicates the Kummer confluent hypergeometric function. 

In this case the relative height of the wave-packet evolves as 
\begin{equation}
\mathcal{V}_{nd}(\varepsilon)=\bigg|{}_1 F_{1} \left[ p_{\rho} , 1; -i\frac{\varepsilon}{\varepsilon_{0}} \left(\frac{1}{f_{\sigma}}-\frac{1}{f_{\rho}}\right)\right]\bigg|^{2}\Theta(\varepsilon).
\end{equation}

In the strongly interacting limit ($\alpha=1$ and consequently $\theta=\pi/4$), due the peculiar functional identities between hypergeometric and the zero-th order Bessel function $J_{0}$,  the above expression reduces to \cite{Degiovanni09}

\begin{equation}
\mathcal{Z}_{nd, strong}(\varepsilon)= \frac{2\pi}{\omega_{c}} e^{i \frac{\varepsilon}{2 \varepsilon_{0}}\left( \frac{1}{f_{\rho}}+\frac{1}{f_{\sigma}}\right)} J_{0}\left[\frac{\varepsilon}{2 \varepsilon_0} \left(\frac{1}{f_{\sigma}}-\frac{1}{f_{\rho}}\right)\right]\Theta(\varepsilon).
\label{Z_strong_int}
\end{equation}
with visibility 
\begin{equation}
\mathcal{V}_{nd, strong}(\varepsilon)=J^{2}_{0}\left[\frac{\varepsilon}{2 \varepsilon_0} \left(\frac{1}{f_{\sigma}}-\frac{1}{f_{\rho}}\right)\right]\Theta(\varepsilon).
\end{equation}

\subsection{Linear dissipation}
The analytic expressions in this case can be obtained from the non-dissipative one by taking into account the substitution 
\begin{equation}
\omega\rightarrow \omega +i \gamma_{1} \omega
\end{equation}
at the level of the first integral. This leads to 
\begin{equation}
\mathcal{Z}_{l}\left(\varepsilon \right)= \frac{2 \pi}{\omega_{c}} e^{i \frac{\varepsilon}{\varepsilon_{0} f_{\rho}}} e^{-\frac{\gamma_{1}}{f_{\rho}}\frac{\varepsilon}{\varepsilon_{0}}}{}_{1} F_{1} \left[p_{\rho}, 1; -\gamma_{1} \frac{\varepsilon}{\varepsilon_{0}}\left(\frac{1}{f_{\sigma}}-\frac{1}{f_{\rho}} \right)+i\frac{\varepsilon}{\varepsilon_{0}}\left(\frac{1}{f_{\sigma}}-\frac{1}{f_{\rho}} \right)  \right]\Theta\left(\varepsilon \right)
\end{equation}
and 
\begin{equation}
\mathcal{V}_{l}\left(\varepsilon \right)= e^{-2 \frac{\gamma_{1}}{f_{\rho}}\frac{\varepsilon}{\varepsilon_{0}}}\bigg|{}_{1} F_{1} \left[p_{\rho}, 1; -\gamma_{1} \frac{\varepsilon}{\varepsilon_{0}}\left(\frac{1}{f_{\sigma}}-\frac{1}{f_{\rho}} \right)+i\frac{\varepsilon}{\varepsilon_{0}}\left(\frac{1}{f_{\sigma}}-\frac{1}{f_{\rho}} \right)  \right]\bigg|^{2}\Theta\left(\varepsilon \right).
\end{equation}

\subsection{Quadratic dissipation}

In this case the elastic scattering amplitude can be written, in the time domain, as
\begin{equation}
\begin{split}
    \mathcal{Z}_{q}(t)&=\mathrm{exp}\left\{\mathcal{W}_{\rho}(t)\right\}\mathrm{exp}\left\{\mathcal{W}_{\sigma}(t)\right\}=\\
    &=\mathrm{exp}\left\{ p_{\rho} \int_{0}^{+\infty} \frac{d\omega}{\omega}[e^{-i\omega(t-\tau_\rho-i\gamma_2\omega\tau_\rho)}-1] e^{-\omega/\omega_c} \right\}\mathrm{exp}\left\{ p_{\sigma} \int_{0}^{+\infty} \frac{d\omega}{\omega}[e^{-i\omega(t-\tau_\sigma-i\gamma_2\omega\tau_\sigma)}-1] e^{-\omega/\omega_c} \right\}.
\end{split}
\end{equation}
This first integration can be done analytically and the exponents $\mathcal{W}_{\rho,\sigma}(t)$ take the following form
\begin{equation}
\begin{split}
    \mathcal{W}_{\rho,\sigma}(t)= 2 p_{\rho,\sigma} \bigg\{ &\gamma-\log(\gamma_2\tau_{\rho,\sigma}\omega_c^2)+i \pi \,\mathrm{Erf}\left[\frac{i+(\tau_{\rho,\sigma}-t)\omega_c}{2\sqrt{\gamma_2\omega_c}}\right]+\\&-\frac{(i+(\tau_{\rho,\sigma}-t)\omega_c)^2}{2\gamma_2\tau_{\rho,\sigma}\omega_c^2} \,{}_2F_2\left[1,1;\frac{3}{2},2;-\frac{(i+(\tau_{\rho,\sigma}-t)\omega_c)^2}{4\gamma_2\tau_{\rho,\sigma}\omega_c^2}\right]\bigg\}
  \end{split}
\end{equation}
where $\gamma\approx 0.577$ is the Euler's constant and  $\mathrm{Erf}$ is the error function. Unfortunately, it is not possible to obtain an analytical solution for the Fourier transform $\mathcal{Z}_q(\varepsilon)$ and a numerical integration is needed.

\section{Comparison with experiments \label{sec:result}}
\begin{figure}[h]
\centering
\includegraphics[scale=0.4]{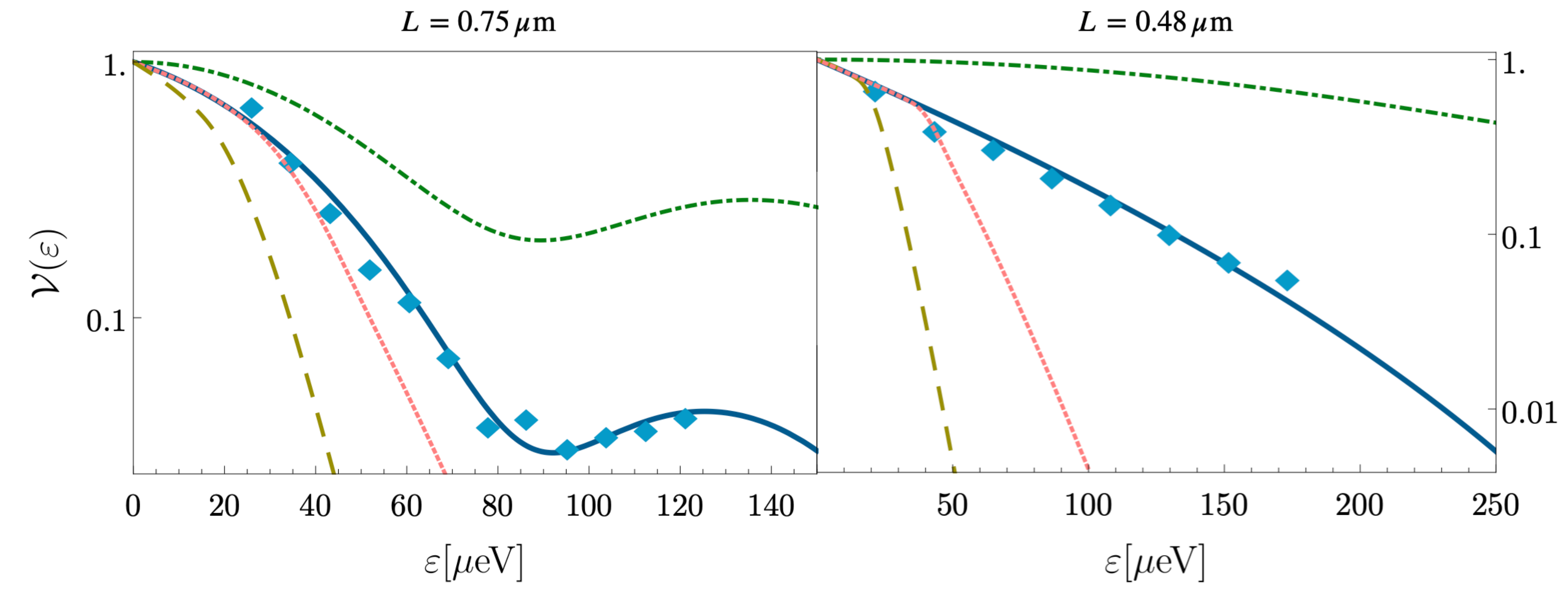}
\caption{Relative peak height as a function of the injection energy (measured in $\mu$eV) for two different sample lengths: $L=0.75\,\mu\mathrm{m}$ (left panel) and $L=0.48\,\mu\mathrm{m}$ (right panel). The non-dissipative case (green dash-dotted curve); the linear dissipative case (blue full curve) with $\gamma_{1}=0.13$ for the left panel and$\gamma_{1}=0.43$ for the right panel; quadratic dissipation with $\gamma_{2}\varepsilon_0=0.03$ for the both panel (red dotted curve) and with $\gamma_{2}\varepsilon_0=0.13$ for the left panel and $\gamma_{2}\varepsilon_0=0.23$ for the right one (bronze dashed curve). Other parameters are: $\alpha=2.1$, $\theta=0.17\pi$ (left panel) and $\alpha=1.6$, $\theta=0.16\pi$ (right panel). Light blue diamonds indicate the experimental data taken from Ref. \cite{Rodriguez20}.}
\label{Fig3}
\end{figure}

The results obtained in the previous Section are shown in Fig.~\ref{Fig3}, where the relative peak height $\mathcal{V}(\varepsilon)$ is plotted versus the injection energy $\varepsilon$ for two different cases compatible to experiments: a sample with length $L=0.75\,\mu\mathrm{m}$ (left panel) and one with $L=0.48\,\mu\mathrm{m}$ (right panel). In both panel, the parameters for the three different dissipative regimes are fixed in order to compare the theoretical expressions with the experimental data (light blue diamonds). In absence of dissipation along the channels (dash-dotted green line) the curve stays above the experimental data due to the absence of exponential overall decay. The observed behaviour is better explained through a linear dissipation model (blue full line). The quadratic dissipation cases considered strongly deviated from the experimental situation because the decay of the relative peak height is more pronounced than the linear one. The discrepancy with the experimental data is more evident for strong dissipation (brown dashed curve), than with weak dissipation (red dotted line). According to these observations, the linear dissipation model can be considered the best candidate to describe the experimental data at least in this case of relatively short propagation lengths ($L<1$ $\mu$m).

It is worth to remark that different experiments \cite{Bocquillon13}, considering a regime of longer propagation lengths ($L>3$ $\mu$m), require to assume a quadratic dissipation to properly reconcile theory and experiments.

\section{Conclusions \label{sec:conclusions}}

In this paper we have investigated the evolution, as a function of the injection energy, of the relative peak height of an electronic wave-packets well resolved in energy and ballistically propagating along a QH edge channels at $\nu=2$. As long as the wave-packet is narrow enough, namely when its width is smaller with respect to the injection energy, this behaviour is well described by the elastic scattering probability of the electronic excitations. In order to be close to experimental observations we have considered a wave-packet crossing an interacting region of variable length where the two edges are capacitively coupled. We have assumed a short range interaction and phenomenologically included in the model a dissipative contribution taking into account the energy dissipation towards external degrees of freedom. According to what discussed in literature, together with the conventional non-dissipative case, we have considered a dissipation both linear and quadratic in the energy. In particular, we have observed that the comparison with the experimental results discussed in Ref. \cite{Rodriguez20} allows us to rule out the non-dissipative case as well a quadratic dependence of the dissipation as a function of the injection energy, and indicates a linear energy loss rate as the more probable candidate to describe the behavior of the wave-packet for these set-ups at short enough lengths ($L< 1$ $\mu$m). This seems in contrast to what discussed in Ref. \cite{Bocquillon13}, where a quadratic dissipation is indicated as dominant contribution in the long propagation length regime ($L>3$ $\mu$m). This discrepancy can be interpreted in two ways: $\emph{i)}$ a strong sample-dependence of the energy dissipation rate; $\emph{ii)}$ a more involved energy dependence of the dissipative contribution on the propagation lengths, not included in our description.

The present analysis has the aim of shading new light on the behavior of electronic wave-packets propagating along ballistic mesoscopic channels and will help both theorists and experimentalists to identify new strategies to mitigate relaxation and dissipative effects that are detrimental for the actual implementation of  flying qubits in solid state devices.

\vspace{6pt}

\appendix
\section{Calculation of the elastic scattering amplitude $\mathcal{Z}$} \label{AppA}

We start by considering an electron injected in channel $1$ in such a way that 

\begin{equation}
 |\mathrm{in}\rangle = \int^{+\infty}_{-\infty} d y \varphi(y) \Psi^{\dagger}(y)|\mathrm{F}\rangle_{1} \otimes |\mathrm{F}\rangle_{2}
\label{in_e} 
\end{equation}
with $|\mathrm{F}\rangle_{i}$ ($i=1,2$) the Fermi sea associated to the $i$-th channel, $\Psi^{\dagger}$ the electronic creation operator and $\varphi(y)$ its wave-packet.

In the following we will focus on an energy resolved wave-packet with
\begin{equation}
\varphi(y)=\frac{e^{i \varepsilon y}}{\sqrt{\mathcal{T}}}
\end{equation}
where the normalization $\mathcal{T}$ represents the longest time scale in the system and $\varepsilon$ the energy. 

According to the hydrodynamic approach discussed in the main text one can write the fermionic operator acting on the Fermi sea as a coherent state of edge-magnetoplasmons (up to a Klein factor that plays no role in what follows) \cite{Wen95}. This leads to 

\begin{equation}
 |\mathrm{in}\rangle = \int^{+\infty}_{-\infty} d y  \frac{e^{i \varepsilon y}}{\sqrt{\mathcal{T}}}\left(\bigotimes_{\omega >0}
 |-\lambda_{\omega}(y)\rangle_{1}\right) \otimes \left( \bigotimes_{\omega>0} |0_{\omega}\rangle_{2} \right)
\label{in_emp} 
\end{equation}
with 
\begin{equation}
\lambda_{\omega}(y)=-\frac{1}{\sqrt{\omega}}e^{i \omega y}
\end{equation}
and $0_{\omega}$ the edge-magnetoplasmon vacuum. 

The analogous expression 
\begin{equation}
 |\mathrm{out}\rangle = \int^{+\infty}_{-\infty} d y'  \frac{e^{i \varepsilon y'}}{\sqrt{\mathcal{T}}}\left(\bigotimes_{\omega >0}
 |-\lambda_{\omega}(y')\rangle_{1}\right) \otimes \left( \bigotimes_{\omega>0} |0_{\omega}\rangle_{2} \right)
\label{out_emp} 
\end{equation}
holds for the state in the outgoing region. 

Expressing the incoming edge-magnetoplasmons in terms of the outgoing ones requires to take into account the entries of the matrix $\hat{S}$ in Eq. (\ref{S_matrix}) in such a way that 
\begin{equation}
|\mathrm{in}\rangle\rightarrow |\mathrm{in}\rangle' = \int^{+\infty}_{-\infty} d y  \frac{e^{i \varepsilon y}}{\sqrt{\mathcal{T}}}\left(\bigotimes_{\omega >0}
 |-S_{11}(\omega)\lambda_{\omega}(y)\rangle_{1}\right) \otimes \left( \bigotimes_{\omega>0} |-S_{12}(\omega)\lambda_{\omega}(y)\rangle_{2} \right).
\label{in_emp} 
\end{equation}

The elastic scattering amplitude is then given by 
\begin{equation}
\mathcal{Z}(\varepsilon)=\langle \mathrm{out}|\mathrm{in}\rangle'
\end{equation}
that, taking into account the general relation for coherent states 
\begin{equation}
\bigotimes_{\omega >0} \langle \alpha_{\omega}|\beta_{\omega}\rangle =e^{-\frac{1}{2} \int^{+\infty}_{0} |\alpha_{\omega}-\beta_{\omega}|^{2}d \omega}e^{i\int^{+\infty}_{0} \Im{\left(\alpha^{*}_{\omega}\beta_{\omega}\right)}d \omega} 
\end{equation}
with $\Im(...)$ representing the imaginary part, leads (in the limit $\mathcal{T}\rightarrow +\infty$) to 
\begin{equation}
\mathcal{Z}\left(\varepsilon \right)= \int_{-\infty}^{+\infty}d\tau e^{i \varepsilon \tau} \exp{\left\{\int_{0}^{+\infty}\frac{d \omega}{\omega} \left[t \left( \omega \right) e^{-i \omega \tau}-1 \right] \right\}}
\end{equation}
which is the expression considered in the main text.

\authorcontributions{Conceptualization, G.R., D.F., R.H.R. and F.P.; formal analysis, G.R., D.F. and F.P.; writing—original draft preparation, G.R. and D.F.; writing—review and editing, R.H.R., F.P., P.R. and M.S.; supervision, P.R. and M.S.}


\acknowledgments{The authors thank N. Traverso Ziani for useful discussions.}

\conflictsofinterest{The authors declare no conflict of interest.} 

\abbreviations{The following abbreviations are used in this manuscript:\\
\noindent 
\begin{tabular}{@{}ll}
QH & Quantum Hall\\
EQO & Electron Quantum Optics\\
\end{tabular}}


\reftitle{References}

\end{document}